\documentclass{llncs}%[a4paper,11pt,twocolumn]

\usepackage{makeidx}  % allows index generation
\usepackage{graphicx} % standard LaTeX graphics tool
\usepackage{subeqnar} % subnumbers individual equations
\usepackage{amsmath}
\usepackage{amssymb}
\usepackage{amscd}
\usepackage{latexsym}
\usepackage{epsfig}
\newtheorem{thm}{Theorem}%[section]

\newtheorem{lem}[thm]{Lemma}

\begin{document}

 %\twocolumn[
 %\begin{@twocolumnfalse}

\title{On Randomized and Quantum Query Complexities}%
\author{Gatis Midrij\=anis}%
%\address{Computer Science Division, U.C. Berkeley}%
%\email{gatis@.cs.berkeley.edu}%
\institute{Computer Science Division \\ U.C. Berkeley \\ E-mail:
gatis@cs.berkeley.edu}

\maketitle

%\thanks{}%
%\subjclass{General.}%
%\keywords{decision trees, randomized computation, quantum computation, lower bounds, degree}%

%\date{}%
%\dedicatory{}%
%\commby{}%
% ----------------------------------------------------------------

\begin{abstract}
We study randomized and quantum query (a.k.a. decision tree)
complexity for all total Boolean functions, with emphasis to
derandomization and dequantization (removing quantumness from
algorithms). Firstly, we show that $D(f) = O(Q_1(f)^3)$ for any
total function $f$, where $D(f)$ is the minimal number of queries
made by a deterministic query algorithm and $Q_1(f)$ is the number
of queries made by any quantum query algorithm (decision tree analog
in quantum case) with one-sided constant error; both algorithms
compute function $f$. Secondly, we show that for all total Boolean
functions $f$ holds $R_0(f)=O(R_2(f)^2 \log N)$, where $R_0(f)$ and
$R_2(f)$ are randomized zero-error (a.k.a Las Vegas) and two-sided
(a.k.a. Monte Carlo) error query complexities.

\end{abstract}
%    \end{@twocolumnfalse}
%  ]

%\maketitle
% ----------------------------------------------------------------

\section{Introduction}\label{sec:intro}

The Boolean query (a.k.a. decision tree) model is probably the
simplest model of a non-uniform computation. In this model, an input
is $N$ bit string $x_1, ..., x_N$ and we want to compute a Boolean
function $f(x_1, ..., x_N)$. The decision tree complexity $D(f)$ is
the minimal number of queries necessary to compute $f(x_1, ...,
x_N)$, where each query asks the value of one variable. It is easy
to see that to query all $x_i$'s is always enough to compute $f$,
thus $D(f) \leq N$ for all $f$. Let us restrict to the case of total
Boolean functions in the rest of the paper.

Randomness, as well as the laws of the quantum world, offer
constructions of new models of computation. In this paper we study
generalizations of the Boolean query model - randomized and quantum
query algorithms. It is known that randomized and quantum query
complexities are polynomially related with the deterministic query
complexity. In this paper our goal is to tighten this result as much
as possible.

Unfortunately, we know only some good randomized query algorithms
\cite{Snir,SaksWig} but surprisingly powerful the query model
appeared in quantum case, since many quantum algorithms are stated
in query model, i.e. Grover's~\cite{Grover} algorithm for OR,
Deutsch-Jozsa's~\cite{DeutschJozsaXOR} algorithm for PARITY and
Ambainis's~\cite{AmbElemDist} element distinctness algorithm.

Like in uniform models, there are at least three variants how much
we allow the randomness in the query model - \emph{zero-error}
(a.k.a. Las Vegas), \emph{one-sided error} and \emph{two-sided
error} (a.k.a. Monte Carlo); let $R_0$, $R_1$ and $R_2$ denote
respective complexities (the optimal number of queries). Similar
situation is in the quantum case also, so let $Q_0$, $Q_1$ and $Q_2$
denote the complexities in quantum case, respectively. It is
interesting that there is meaningful to define \emph{exact}
complexity ($Q_E$) in the quantum model. Accurate definitions we
will give in Section~\ref{sec:prelim}.

There are many ways how those complexities can be compared. Two of
them are the most popular ones. The first, to compare them for some
particular function (or some class of functions). Over the last
years, a rich body of work has been investigated to show both upper
and lower bounds of certain functions both for randomized complexity
(i.e. \cite{InfCompl,SaksWig,Snir}) and quantum complexity (i.e.
\cite{AaronsonColl,AmbCol,Kutin,ShiColl,BuhrZeroBounds,Grover}).

The second way, to which we focus in this paper, is to show
relations betweens those models that hold for \emph{all} functions.
Other ways include studying the complexity of random functions and
an average-case complexity.

The next section briefly survives known results and states our ones,
as well as gives organization of the paper. Note: in the rest of the
paper, unless otherwise specified, all results hold for every total
Boolean function $f$.

\section{Previous work and our results}\label{sec:prev}
\subsection{The random case}\label{sec:prev_r}

Trivially, $R_2(f) \leq R_1(f) \leq R_0(f) \leq D(f)$. The first
non-trivial result follows from an independent work of several
authors \cite{BlumImp,HartmHem,Tardos} and states that $D(f) =
O(R_0(f)^2)$. Nisan~\cite{NisanCrew} generalized it to one-sided
error case $D(f) = O(R_1(f)^2)$ and to two-sided error case $D(f) =
O(R_2(f)^3)$. In this paper we show that $R_0(f) = O(R_2(f)^2 \log
N)$, where $N$ is the length of input.

Much more progress has been made to study the complexity for certain
classes of functions. For instance, it is known that $D(g) =
O(R_0(g)^{1.96...})$ \cite{HeimNwWig,HeimWig} holds for every
read-once formula $g$. Santha~\cite{SanthaRead} showed that $R_0(g)
= \Theta(R_2(g))$ holds for the class of read-once formulas $g$. For
other classes of functions, like graph properties, monotone
functions, random functions and symmetric functions, better results
are known, too.

The best randomized algorithm is given by Snir~\cite{Snir}; he shows
that for the recursive NAND function, $D(NAND) =
\Omega(R_0(NAND)^{1.326...})$. For NAND function this algorithm is
tight \cite{SaksWig} and this gap is conjectured to be an optimal
separation between deterministic and randomized complexities.

\subsection{The quantum case}\label{sec:prev_q}

Trivially, $Q_2(f) \leq Q_1(f) \leq Q_0(f) \leq Q_E(f)$ and $Q_2(f)
\leq R_2(f)$, $Q_1(f) \leq R_1(f)$, $Q_0(f) \leq R_0(f)$, $Q_E(f)
\leq D(f)$. Beals et al. \cite{Beals} showed that $D(f) =
O(Q_2(f)^6)$ and $D(f) = O(Q_E(f)^4)$. Buhrman et al.
\cite{BuhrZeroBounds} improved the later to $D(f) = O(Q_1(f)^4)$
\footnote{Explicitly they showed it only for $Q_0$ but it is easy to
generalize to $Q_1$ case.}. Aaronson~\cite{AaronsonCert} showed a
relation between one-sided error randomized and quantum
complexities, $R_0(f) = O(Q_1(f)^3\log N)$ \footnote{The same note
as in the previous footnote.}. We give better result, $D(f) =
O(Q_1(f)^3)$.

Again, as well as in the random case, none of those relations are
believed to be tight. Quantum algorithms usually are much more
sophisticated than randomized ones. Countless papers have been
written to find fast quantum algorithms as well as to characterize
the power of quantum lower bound techniques. The best known quantum
query algorithm is Grover's algorithm \cite{Grover} for OR function
that gives $R_2(OR) = \Omega(Q_1(OR)^2)$ and $Q_0(OR) =
\Omega(Q_1(OR)^2)$ \cite{Beals}. Buhrman et al.
\cite{BuhrZeroBounds} showed that for any $\varepsilon > 0$ there is
a function $g_{\varepsilon}$ such that $R_2(g_{\varepsilon}) =
\Omega(Q_0(g_{\varepsilon})^{2-\varepsilon})$ and
$Q_E(g_{\varepsilon}) =
\Omega(Q_0(g_{\varepsilon})^{2-\varepsilon})$ \footnote{Actually
they showed only the first case but the second case is a trivial
application of polynomials method.}. The best known separation
between $Q_E(f)$ and $D(f)$ is just by a factor $2$
\cite{DeutschJozsaXOR}. The result by van Dam~\cite{Dam} shows that
$Q_2(f) \leq \frac{N}{2}+\sqrt{N}$.

\subsection{The organization of the paper}\label{sec:prev_org}
The rest of the paper is organized as follows.
Section~\ref{sec:prelim} gives definitions and some basic results we
will use in proofs. Section~\ref{sec:d_vs_q1} proves the relation
between deterministic and quantum complexities.
Section~\ref{sec:r0_vs_r2q2} proves the relation between randomized
complexities. At the end, section~\ref{sec:concl} gives some
immediate extensions of the results in this paper.

\section{Preliminaries}\label{sec:prelim}

We assume familiarity with classical and quantum query algorithms
and basic complexity measures of them, so we will quickly breeze
through definitions, notation and basic results. For more explicit
statement one can look in superb (but somewhat outdated) survey by
Buhrman and de Wolf \cite{BuhrWolfSurvey}; mostly this section is
based on the work done by Nisan~\cite{NisanCrew}, Beals et al.
\cite{Beals} and de Wolf~\cite{WolfND}.

We consider computing a Boolean function $f(x_1,...,x_N):\{0,1\}^N
\rightarrow \{0,1\}$ in the query model. In this model, the input
bits can be accessed by queries to an oracle $X$ and the complexity
of $f$ is the number of queries needed to compute $f$. The
deterministic query complexity $D(f)$ is just a minimal number of
queries necessary to compute function $f$.

A randomized query algorithm is just a probability distribution over
deterministic query algorithms. We are interested in algorithms
making minimal number of queries in the worst-case such that  for
all inputs it returns correct answer with probability at least $\rho
\geq 4/5$ \footnote{$4/5$ can be replaced by arbitrary constant in
(1/2..1).}.

A quantum computation with $T$ queries is just a sequence of unitary
transformations
$$U_1 \rightarrow O \rightarrow ...
\rightarrow U_{T-1} \rightarrow O \rightarrow U_{T} \rightarrow O.$$

$U_j$ can be arbitrary unitary transformation that do not depend on
the input bits $x_1,...,x_N$. $O$ are query transformations. To
define $O$, we represent basis states as $| i, z \rangle$ where $i$
consists of $\lceil logN \rceil$ bits describing an index of a
variable to be queried, $b$ is one bit and $z$ consists of all other
bits. Then, $O$ maps $|i, b, z \rangle$ to $(-1)^{b·x_i} |i, b, z
\rangle$ (i.e., we change phase depending on $x_i$). The computation
starts with a state $|0 \rangle$. Then, we apply $U_1, O, . . ., O,
U_T$ and measure the final state. The result of the computation is
the bit $b$ obtained by the measurement. Now we can define the
models depending on probability $\rho$ such that for every $x =
(x_1, ..., x_N)$, the rightmost bit of $U_TO_x . . .O_xU_1|0
\rangle$ equals $f(x_1, . . . , x_N)$ with probability at least
$\rho = \rho(x_1,...,x_N)$.

Both in randomized and quantum algorithms we are interested in those
ones who compute functions asking as less queries as possible. With
complexity of an algorithm we mean the number of queries it make.
For such algorithms, if $\rho=1$ then randomized query complexity is
equal with deterministic complexity one but quantum complexity is
denoted by $Q_E(f)$. If $\rho=1$ on all $0$-instances or all
$1$-instances and always $\rho \geq 4/5$ then we call it one-sided
error algorithm and $R_1(f)$ ($Q_1(1))$ denote respective
complexities. If $\rho \geq 4/5$ then we call the complexity
two-sided and denote $R_2(f)$ and $Q_2(f)$, respectively. Zero-error
case is special, because algorithms are allowed to output also "?"
(meaning "I don't know"). When it outputs $0$ or $1$ then it should
be correct always but it can output $?$ with probability at most
$1/5$. Let $R_0(f)$ and $Q_0(f)$ denote the corresponding
complexities.

It is well known fact that a Boolean function is unique represented
by a multilinear polynomial. For example, $OR$ function is
represented by a polynomial $1-(1-x_1)(1-x_2)...(1-x_N)$.
Polynomials that approximate functions on every input will be
interesting too.

Beals et al. showed the source lemma for polynomial method:
\begin{lem}\cite{Beals}
The probability to output correct answer for every quantum query
algorithm making $T$ queries is described by a multilinear
polynomial with degree at most $2T$.
\end{lem}
{\em Sketch of the proof.} The state of an algorithm running on
input $x$ can be described by $$\sum\limits_{i,b,z}
p_{i,b,z}(x)|i,b,z>$$ where $p_k$ is an amplitude of basis state
$k$. At the beginning $p_{i,b,z}(x)$ do not depends on input word
$x$. After a query, amplitudes changes. Since oracle is a linear
operator, we can analyze each monomial separately. Oracle maps
$p_{i,b,z}(x)|i,b,z>$ to $(-1)^{b·x_i}p_{i,b,z}(x)|i,b,z>$. Since
$(-1)^y=(1-2y)$ then $(-1)^{b·x_i}p_{i,b,z}(x)|i,b,z>$ = $(1 -
2b·x_i)p_{i,b,z}(x)|i,b,z>$. So the degree of amplitudes increase
just by at most one.

Unitaries cannot increase the degree. Measurement can at most double
the degree because the probability to observe a state to be $0$ on
$b$ is
$$\sum\limits_{i,0,z} |p_{i,0,z}(x)|^2.$$\qed

This allows us to lower bound the number of queries of quantum
algorithm by finding lower bound of degree by polynomial
representing function. More precisely, $Q_E(f) \geq deg(f)/2$,
$Q_0\geq ndeg(f)/2$, $Q_1\geq min(ndeg(f),ndeg(1-f))/2$ and $Q_2\geq
\widetilde{deg}(f)/2$. There $deg(f)$ denotes the degree of
polynomial representing $f$, $ndeg(f)$ (degree of a nondeterministic
polynomial) denotes the minimal degree of polynomial $p$ such that
$p(x)=0$ iff $f(x)=0$, $\widetilde{deg}(f)$ denotes the minimal
degree of polynomial approximating function $f$ on every input. It
is easy to see that $deg(f) \geq ndeg(f) \geq \widetilde{deg}(f)$.

The block sensitivity of $f$ on $x$ is the maximum number of
disjoint $B_j\subseteq \{1, \ldots, n\}$ such that $f(x^{B_j})\neq
f(x)$, $x^{B_j}$ being $x$ with all $x_i$ for $i\in B_j$ changed to
$1-x_i$. We denote it $bs_x(f)$. Let $bs(f)=\max bs_x(f)$. The
sensitivity $s(f)$ is the same just all blocks are restricted to be
with a size one. It is easy to see that $s(f) \leq bs(f)$.

It is known that
\begin{thm}\label{thm:BlockSensVSDegree} \cite{NisanSzegedy}
For any total Boolean function $f$, $$bs(f) =
O(\widetilde{deg}(f)^2).$$
\end{thm}
\begin{thm}\label{thm:RvsBS} \cite{NisanCrew}
For any Boolean function $f$, $$R_2(f) \geq bs(f)/2.$$
\end{thm}
\proof Let $w$ be the input that achieves the block sensitivity, and
let $B_1, B_2,...B_t$ be the disjoint sets s.t. $f$ is sensitive to
$B_i$ on $w$. For each $1 \leq i \leq t$, any randomized algorithm
running on $w$ must query some variable in $B_i$ with probability of
at least 1/2, since otherwise it cannot distinguish between $w$ and
$w^{B_i}$. Thus the total number of queries has to be at least
$t/2$.\qed

\section{Deterministic vs. quantum one-sided error}\label{sec:d_vs_q1}

To dequantize one-sided error algorithms we use polynomials method.
Our result is improvement over that ones by Buhrman et al.
\cite{BuhrZeroBounds} and Aaronson~\cite{AaronsonCert}. Here
\emph{maxonomial} of polynomial $p$ is a monomial with maximal
degree.

The following generalization of a lemma attributed in \cite{Beals}
to Nisan and Smolensky was independently observed by
Aaronson~\cite{AaronsonCert}. The key idea of it is that querying a
maxonomial, we decrease the function's block sensitivity on any
input word by at least one.

\begin{lem}\label{lemma:maxonBS}
    For any nondeterministic polynomial $p$ approximating function $f$,
    for every $0$-instance $w \in \{0,1\}^N$ (s.t. $f(w) = 0$)
    and every maxonomial $M$ of
    $p$, there is a set $B$ of variables in $M$ such that $f(w^B)  =1 $.
\end{lem}
\proof Obtain restricted polynomial $g$ from $p$ by setting all
variables outside of $M$ according to $w$. Obtain word $w' \in
\{0,1\}^{|M|}$ that assigns values from $w$ to variables in $M$.
Since $g$ makes no errors on $0$-instance, $g(w') = 0$. This $g$
contains monomial $M$ therefore it cannot be constant $0$. Therefore
there is some set $B$ of variables in $M$ that makes $g(w'^B)
> 0$ and hence $f(w^B) = 1$.

\qed

This we use in the following algorithm.

\begin{lem}\label{thm:detVSdegree}
    For every total Boolean function $f$,
     $$D(f) \leq (bs(f)+1)*ndeg(f).$$
\end{lem}
\proof

The deterministic query algorithm $\mathcal{A}$ is written in pseudo
code, as a function of a complete description of a polynomial $q$
that nondeterministically represent the function $f$
\footnote{Remember, it means that for every input word $x$, $q(x)=0$
if and only if $f(x)=0$} (thus $deg(q) = ndeg(f)$) and a word $X \in
\{0,1\}^N$ given by queries. $\mathcal{A}$ returns value of $f(X)$.
A function $sign: \mathcal{R} \rightarrow \{0,1\}$ is defined as
follows; if $p \neq 0$ let $sign(p)=1$ otherwise let $sign(p)=0$.
The algorithm $\mathcal{A}$:

\begin{minipage}[h]{12 cm}
 \begin{description}
    \item[] $\{0,1\}$ function Value$\diamond$f\{\\By value q as polynomial, \\ By queries $X \in
    \{0,1\}^N$;
        \begin{description}
            \item[1] $p := q$;
            \item[2] Repeat $bs(f)+1$ times \{
            \begin{description}
                \item[3] If $p$ is constant then return $sign(p)$;
                \item[4] Pick a maxonomial $M$ in $p$;
                \item[5] Query X-values of $M$'s variables;
                \item[6] Replace all queried variables in $p$ \\ by appropriate
                constants;
            \end{description}
            \item[] \};
            \item[7] Return $1$;
        \end{description}
    \item[] \};
 \end{description}
\end{minipage}

The nondeterministic "pick a maxonomial" can easily be made
deterministic by choosing the the first maxonomial in some fixed
order.

It is easy to see that for every maxonomial $M$ holds $|M| = deg(p)$
and at every moment $deg(p) \leq deg(q)$, thus in every cycle
$\mathcal{A}$ makes at most $deg(q)$ queries, hence the number of
queries $\leq deg(q)*bs(f)$. If $\mathcal{A}$ returns the answer in
3rd line then it is right because $q$ represents $f$. If input word
is 0-instance then by Lemma~\ref{lemma:maxonBS}, querying each
maxonomial decreases the function's block sensitivity on $x$; after
$bs(f)$ repetitions it should be a constant. Therefore algorithm can
reach 7th line only on 1-instances.

%We will do it by proving the following statement: if cycle is
%executed $a$ times then the function's $f$ block sensitivity on the
%word $X$ is at least $a$, thus $a \leq bs_X(f) \leq bs(f)$.
%
%By induction we will show the statement: $bs_X(f) \geq a$ and
%variables in blocks can be taken only from yet queried variables.
%Bases: when $a=0$, it is obvious. Inductive assumption: after $a-1$
%executions of cycle, $X$ has at least $a-1$ disjoint blocks that
%take their variables only from yet queried variables and to which
%$f$ is sensitive on $X$. We will show that in the next cycle's
%execution there appears a block $B$ that takes its variables only
%from variables queried in this new cycle (therefore is disjoint with
%previous ones) and to which $f$ is sensitive to $X$.
%
%Let $M$ denote maxonomial chosen in this cycle. Let $w \in
%\{0,1\}^{|M|}$ denote the word that takes its variables from $M$ and
%to whom $p(w)$ is a nondeterministic polynomial of $f_X(w)$, where
%$f_X(w)$ denotes the function $f$ which is restricted outside domain
%$M$ according to $X$. Such exists, since $p$ is just polynomial $q$
%where some variables are replaced with constants according to $X$.
%It is easy to see that for any set of variables $B \subseteq M$
%holds $f_X(w^B) = f(X^B)$. Now Lemma~\ref{lemma:maxonBS} says that
%there is a set $B$ of variables in $M$ such that $f_X(w^B) = 1$.
%Since $f(X) = 0$ and $f(X^B) = f_X(w^B) = 1$ it follows that $f(X)
%\neq f(X^B)$. This concludes induction.

\qed

\begin{thm}\label{thm:quant_e}
    For every total Boolean function $f$,
     $$D(f) = O(Q_1(f)^3).$$
\end{thm}
\proof Lemma~\ref{thm:detVSdegree} and
Theorem~\ref{thm:BlockSensVSDegree} gives this relation whenever
quantum algorithm happens to make error on $1$-instances. However,
if it makes error on $0$-instances we could just dequantizate the
complementary function, and afterward just flip all the answers in
the deterministic decision tree \footnote{Notice, that in general a
statement "$Q_1(f) \geq ndeg(f)/2$" is not true but in our case we
have a relation that is true for \emph{all} functions $f$ therefore
also for complementary functions.} .\qed

\section{Randomized zero-error vs. random two-sided error}\label{sec:r0_vs_r2q2}

Before this paper, the only nontrivial \footnote{Special case of
Nisan's result - $R_0(f)=O(R_2(f)^3)$ I call trivial.} relation
between $R_0$ and $R_2$ was $$R_0(f)=O(R_2(f)ndeg(f)\log N)$$ by
Aaronson~\cite{AaronsonCert}. In this section we prove
\begin{thm}\label{thm:r0VSr2} For every total Boolean function $f$,
$$R_0(f) = O(R_2(f)^2\log N).$$
\end{thm}

Nisan introduced minimal sensitive blocks on input word $X$ as
sensitive blocks whom any strict subset is not sensitive on $X$ and
proved
\begin{lem}\cite{NisanCrew}
For every word $X$, for every minimal sensitive block $B$ on $X$,
$$|B| \leq s(f) \leq bs(f).$$
\end{lem}
\proof If we flip one of the B-variables in $X^B$, then the
function's value must flip as well (otherwise $B$ would not be
minimal), so every B-variable is sensitive for $f$ on $X^B$.\qed

We can easily get very rough estimate of the number of minimal
sensitive blocks for $f$ on word $X$:
\begin{lem}\label{lem:numbBlocks} For any total Boolean function
$f$ and word $X$, the number of minimal sensitive blocks on word $X$
is at most $N^{bs(f)}.$
\end{lem}
\proof Since the previous lemma said that the size of any minimal
sensitive block cannot be bigger than $bs(f)$, then the maximal
number of minimal sensitive blocks is less than the number of
subsets of at most $bs(f)$ variables, that is
$\sum\limits_{k=1}^{bs(f)}\binom{N}{k} \leq N^{bs(f)}$.
   \qed

The proof of the Theorem~\ref{thm:r0VSr2} just follows from the next
lemma, by applying the Lemma~\ref{thm:RvsBS}.

\begin{lem}\label{lem:r0VSr2bs} For every total Boolean function $f$,
$$R_0(f) = O(R_2(f)bs(f)\log N).$$
\end{lem}
\proof The zero-error randomized algorithm running on word $X$ is as
follows. Repeat two-sided error algorithm and take majority, until
it gives estimation of expected error $\epsilon \leq
\frac{1}{2N^{bs(f)}}$. We need $\Theta(bs(f)\log N)$ repetitions to
get it (as usual, by Chenoff's bounds). If at this moment the value
of $f(X)$ is not determined for sure \footnote{For those who know
the notion of a "certificate". We output "?" if we have not found a
certificate.} then output "?"; otherwise output the value. To finish
the proof we have to show that the value of $f(X)$ is determined
\footnote{In other words, we have found a certificate.} with
probability at least $1/2$.

Assume not; then there exists a block $B \subseteq [N]$ such that
$f(X) \neq f(X^B)$, moreover, there should be such minimal block. On
the other hand, by a simple adversary argument, every sensitive
block $B$ of function $f$ on word $X$ should be queried with
probability at least $\epsilon$. The expected number of blocks which
are not touched is at most $\epsilon*N^{bs(f)} \leq 1/2$ (by the
Lemma~\ref{lem:numbBlocks}). Therefore with probability at least
$1/2$ there are no minimal sensitive blocks left, thus the value of
$f(X)$ is determined.

\qed

\section{Extension of results}\label{sec:concl}

In the previous sections, to make picture simpler we compared just
two complexities in each inequality. Actually, one could wish to see
those results more precisely. Now we review all of them. All
inequalities in the list hold for every total function $f$ up to
constant factor:
\begin{itemize}
    \item $D(f) \leq R_0(f)^2$ \cite{BlumImp,HartmHem,Tardos}.
    \item $D(f) \leq R_1(f)R_2(f)$ \cite{NisanCrew}.
    \item $D(f) \leq R_2(f)^3$ \cite{NisanCrew}.
    \item $R_0(f) \leq R_2(f)^2 \log N$ [Theorem~\ref{thm:r0VSr2}].
    \item $D(f) \leq Q_2(f)^6$ \cite{Beals}.
    \item $D(f) \leq Q_E(f)^2Q_2(f)^2$ \cite{Beals}.
    \item $D(f) \leq Q_1(f)^2Q_2(f)^2$ \cite{BuhrZeroBounds}.
    \item $R_0(f) \leq Q_1(f)Q_2(f)^2 \log N$
    \cite{AaronsonCert}.
    \item $D(f) \leq Q_1(f)Q_2(f)^2$ [Lemma~\ref{thm:detVSdegree}].
\end{itemize}

Probably non of those inequalities are tight.

%The other way to extend our results is average-case setting (versus
%worst-case setting we used before). Average-case complexity concerns
%the expected number of queries needed to compute some function $f$
%when the input is distributed according to some given probability
%distribution $\mu$. Let $C^\mu(f)$ denote the average-case analogue
%of the worst-case complexity $C \in \{D, R_0, R_1, R_2, Q_E, Q_0,
%Q_1, Q_2\}$.

\section*{Acknowledgements}
I would like to thank Scott Aaronson about many useful comments
about the paper. I thank Ronald de Wolf for noticing that the
algorithm in section ~\ref{sec:d_vs_q1} gives lower bound not only
for exact quantum algorithms.

% ----------------------------------------------------------------
\bibliographystyle{amsplain}

\end{document}